\documentclass[aps,prd,twocolumn,,nofootinbib]{revtex4-1}
\usepackage{epsfig}
\usepackage{color}

\begin{document}
\author{Pei-lin Yin$^{1}$, Yuan-mei Shi$^{2,3}$, Zhu-fang Cui$^{2,6}$, Hong-tao Feng$^{4}$, and Hong-shi Zong$^{2,5,6,}$}\email[]{zonghs@nju.edu.cn}
\address{$^{1}$Key Laboratory of Modern Acoustics, MOE, Institute of Acoustics, and Department of Physics, Nanjing University, Nanjing 210093, China}
\address{$^{2}$Department of Physics, Nanjing University, Nanjing 210093, China}
\address{$^{3}$Department of Physics and electronic engineering, Nanjing Xiaozhuang University, Nanjing 211171, China}
\address{$^{4}$Department of Physics, Southeast University, Nanjing 211189, China}
\address{$^{5}$Joint Center for Particle, Nuclear Physics and Cosmology, Nanjing 210093, China}
\address{$^{6}$State Key Laboratory of Theoretical Physics, Institute of Theoretical Physics, CAS, Beijing, 100190, China}

\title{Continuum study of various susceptibilities within thermal QED$_3$}
\begin{abstract}
In this paper, the relations of four different susceptibilities (i.e., the chiral susceptibility, the fermion number susceptibility, the thermal susceptibility and the staggered spin susceptibility) are investigated both in and beyond the chiral limit. To this end, we numerically solve the finite temperature version of the truncated Dyson-Schwinger equations for fermion and boson propagator. It is found that, in the chiral limit, the four susceptibilities give the same critical temperature and signal a typical second order phase transition. But the situation changes beyond the chiral limit: the critical temperatures from the chiral and the thermal susceptibilities are different, which shows that to define a critical region instead of an exclusive point for crossover might be a more suitable choice; meanwhile, both the fermion number and the staggered spin susceptibilities have no singular behaviors any more, this may mean that they are no longer available to describe the crossover properties of the system.
\bigskip

\noindent PACS Numbers:  11.10.Kk, 11.15.Tk, 11.30.Qc
\end{abstract}
\maketitle

\section{INTRODUCTION}
The Quantum Chromodynamics (QCD) vacuum exhibits many non-perturbative phenomena that don't present in Quantum Electrodynamics (QED) vacuum. As the temperature and/or chemical potential increase, the QCD vacuum changes and the system undergoes a phase transition into another phase when the temperature and/or chemical potential reach its critical values. There may be many different phases corresponding to different regions of temperature and/or chemical potential. The researches for properties of strongly interacting matter within different phases as well as behaviors of phase transitions, so as to map the phase diagram for the system in the plane of temperature and chemical potential, are quite important in today's basic physics theory and experiment, so a great many of studies has been done on this field~\cite{Phys.Rev.D.58.096007,Phys.Rev.Lett.86.592,Nucl.Phys.B.673.170,Nature.443.675,Rev.Mod.Phys.80.1455,Rep.Prog.Phys.74.014001,J.HighEnergyPhys.04.014,Eur.Phys.J.C.73.2612}.

In drawing the phase diagram of strongly interacting matter, the researches of chiral symmetry breaking-restoration and confinement-deconfinement phase transitions are important aspect. Some quantities that characterize the above two kinds of phase transition are introduced both in Lattice QCD simulation and continuum model studies, for instance, the chiral fermion condensate, the chiral susceptibility~\cite{Phys.Rev.D.50.6954,Phys.Lett.B.591.277,Phys.Lett.B.643.46,
Phys.Rev.C.79.035209}, the quark number susceptibility~\cite{Phys.Rev.Lett.59.2247,Phys.Rev.Lett.91.102003,Phys.Rev.D.67.014028,Phys.Rev.D.75.074013,
Phys.Rev.D.82.054026,J.HighEnergyPhys.02.066}, the thermal susceptibility~\cite{Phys.Rev.C.58.1758,Phys.Rev.D.84.074020,Phys.Rev.D.85.034026} and so on. For the latter, the approach includes chiral perturbation theory~\cite{Phys.Rev.D.54.1087}, Dyson-Schwinger equations (DSEs) and Bethe-Salpeter equation (BSE)~\cite{Phys.Lett.B.639.248,Phys.Rev.D.80.074029}, hard-thermal/dense-loop (HTL/HDL) approximation~\cite{Phys.Lett.B.523.143,Phys.Rev.D.68.085012,J.Phys.G:Nucl.Part.Phys.37.055001}, Nambu-Jona-Lasinio (NJL) model~\cite{Nucl.Phys.A.576.525,Phys.Rev.D.77.114028,Phys.Rev.D.88.114019}, QCD sum rule and others. With regard to chiral phase transition, the chiral and the quark number susceptibilities have attracted many interests over the past few years. In Ref.~\cite{Phys.Lett.B.675.32}, based on a continuum model study of the chiral susceptibility in two-flavor QCD, the authors show that, with temperature increasing, this susceptibility exhibits a divergent peak (which indicates a second order phase transition) in the chiral limit, and a finite peak (which means a crossover) beyond the chiral limit.

Due to the complex non-Abelian character of QCD, it is difficult to have a thorough understanding of chiral phase transition. In this case, to gain valuable comprehension of it, it is very suggestive to study some model, which is structurally much simpler than QCD while sharing the same basic non-perturbative phenomena.

As a field theoretical model, Quantum Electrodynamics in (2+1) dimensions (QED$_3$) has been studied quite intensively in recent years. It has many features similar to QCD, such as dynamical chiral symmetry breaking (DCSB)~\cite{Phys.Rev.Lett.60.2575,Phys.Rev.D.44.540,Phys.Rev.D.54.4049,Phys.Lett.B.491.280,Phys.Rev.D.70.073007,Phys.Rev.C.78.055201,Phys.Rev.D.86.065002} and confinement~\cite{Phys.Rev.D.46.2695,Phys.Rev.D.52.6087,Phys.Rev.Lett.91.171601,Phys.Rev.D.81.034030}. In addition, due to the coupling constant being dimensionful (its dimension
is $\sqrt{mass}$), QED$_3$ is superrenormalizable, so it does not suffer from the ultraviolet divergence which is present in QED$_4$. Apart from these interesting features, QED$_3$ with $\textit{N}_{f}$ massless fermion flavors can be regarded as a possible low energy effective theory for strongly correlated electronic systems~\cite{Phys.Rev.B.66.054535,Nature.438.197,Int.J.Mod.Phys.B.21.4611}.

In order to see what will happen in the case of QED$_3$, the chiral phase transition driven by the temperature in QED$_3$ is investigated by analyzing the temperature dependence of susceptibility in the present paper. Although there are some works on studying the chiral phase transition by susceptibility in the past few years, such as the chiral susceptibility, the fermion number susceptibility and so on, the research of the relations among them in the same framework seems scant, as far as the present authors know. The motivation of present paper is to discuss specifically the behavior of four different susceptibilities (i.e., the chiral susceptibility, the fermion number susceptibility, the thermal susceptibility and the staggered spin susceptibility~\cite{PhysRevD.88.125022}) with the temperature varying in and beyond the chiral limit separately to compare the similarity and difference among them.

This paper is organized as follows: In Sec. II, model-independent analytical expressions for the four susceptibilities are given, which express susceptibilities as integrals of dressed propagators and dressed vertex. In Sec. III, calculations of the four susceptibilities within the DSEs framework are performed. A brief summary and discussions are given in Sec. IV.

\section{ANALYTICAL TREATMENT}
Dynamical properties of a many-particle system can be investigated by measuring the response of the system to an external perturbation that disturbs the system only slightly in its equilibrium state. A noticeable measure is the susceptibilities that are defined as the first-order derivative of the order parameter with respect to the external field. The order parameter is radically different in two phases and thus characterizes the phase transition of the system. As a result, the divergent or some other singular behaviors of susceptibilities are usually regarded as essential characteristics of phase transition.

In this section, by means of the external field method in Ref.~\cite{Phys.Rev.C.72.035202}, model-independent expressions for the four susceptibilities are given.

\subsection{Formalism of the chiral susceptibility}
It is commonly accepted that with the temperature and/or chemical potential increasing, strongly interacting matter will undergo a phase transition from the Nambu-Goldstone phase (or Nambu phase, in which the condensate of particle-antiparticle pairs, the order parameter of the chiral phase transition, is non-zero due to DCSB) to the Wigner phase (where chiral symmetry is partially restored and thus the condensate vanishes). The fluctuation of this order parameter is related to the chiral susceptibility which measures the response of the chiral condensate to a small perturbation of the current mass of the fermion.

The chiral susceptibility is defined as
\begin{eqnarray}
\chi^{c}=\frac{\partial(-\langle\bar\psi\psi\rangle)}{\partial\textit{m}}=\frac{T}{V}\frac{\partial^{2}\ln\mathcal{Z}}{\partial m^{2}},\label{eq1}
\end{eqnarray}
where $\langle\bar\psi\psi\rangle$ is the fermion chiral condensate in the presence of current mass \textit{m}, and $\mathcal{Z}$ denotes the partition function of the system.

Formally, we can express the chiral condensate by means of the dressed fermion propagator
\begin{eqnarray}
-\langle\bar\psi\psi\rangle=\int\frac{\textrm{d}^3p}{(2\pi)^3}\textrm{Tr}[S(\textit{m},p)],\label{eq2}
\end{eqnarray}
where the notation Tr denotes trace over Dirac indices of the propagator, and \textit{S} is the dressed fermion propagator at finite current mass \textit{m}.

Substituting Eq. (2) into Eq. (1) and adopting the identity
\begin{eqnarray}
\frac{\partial S(m,p)}{\partial m}=-S(m,p)\frac{\partial S^{-1}(m,p)}{\partial m}S(m,p),\label{eq3}
\end{eqnarray}
we immediately arrive at
\begin{eqnarray}
\chi^{c}=-\int\frac{\textrm{d}^{3}p}{(2\pi)^{3}}\textrm{Tr}[S(m,p)\frac{\partial S^{-1}(m,p)}{\partial m}S(m,p)],\label{eq4}
\end{eqnarray}

Analogizing the well-known Ward identity in QED, we consider the current mass \textit{m} as a constant background scalar field coupled to the fermion fields by the term $m\bar\psi\psi$. Then \textit{S}(\textit{m},\textit{p}) is the dressed fermion propagator in the presence of such a background field and the derivative of its inverse with respect to \textit{m} yields the so-called one-particle-irreducible (1PI) dressed scalar vertex
\begin{eqnarray}
\Gamma(m,0,p)=\frac{\partial S^{-1}(m,p)}{\partial m},\label{eq5}
\end{eqnarray}
where $\textit{p}$ is the relative momentum, and the total momentum of the dressed scalar vertex vanishes because the background scalar field \textit{m} is a coordinate-independent constant.

Substituting Eq. (5) into Eq. (4) gives
\begin{eqnarray}
\chi^{c}=-\int\frac{\textrm{d}^{3}p}{(2\pi)^{3}}\textrm{Tr}[S(m,p)\Gamma(m,0,p)S(m,p)],\label{eq6}
\end{eqnarray}
therefore, we obtain an integral formula for the chiral susceptibility at zero temperature and chemical potential. It expresses the chiral susceptibility in terms of the dressed fermion propagator and the dressed scalar vertex, which are just basic quantities in quantum field theory. The DSEs-BSE approach provides us a desirable framework to calculate them, and hence the chiral susceptibility.

Here, we note that there is a linear divergence in the above integral. In order to obtain something meaningful from the chiral susceptibility, we need to subtract the linear divergence of the free chiral susceptibility. The regularized chiral susceptibility is defined by
\begin{eqnarray}
\chi^{c}_{r}&=&\chi^{c}-\chi^{c}_{f}\nonumber\\
&=&-\int\frac{\textrm{d}^{3}p}{(2\pi)^{3}}\textrm{Tr}[S(m,p)\Gamma(m,0,p)S(m,p)\nonumber\\
&&-S_{0}(m,p)\textbf{1}S_{0}(m,p)],\label{eq7}
\end{eqnarray}

This expression can be generalized to the case of finite temperature. According to finite temperature field theory, the corresponding finite temperature version of the chiral susceptibility can be obtained by replacing the integration over the third component of the momentum with summation over Matsubara frequencies
\begin{eqnarray}
\chi^{c}_{r}(T)&=&-\textit{T}\sum_{n}\int\frac{\textrm{d}^{2}P}{(2\pi)^{2}}\textrm{Tr}[S(m,p_{n})\Gamma(m,0,p_{n})S(m,p_{n})\nonumber\\
&&-S_{0}(m,p_{n})\textbf{1}S_{0}(m,p_{n})],\label{eq8}
\end{eqnarray}
where $p^{\mu}_{n}=(\omega_{n},\vec{p})$. The notation $\omega_{n}$ denotes the fermion Matsubara frequencies, i.e. $\omega_{n}=(2\textit{n}+1)\pi\textit{T}$, $\vec{p}$ represents the spatial component of the momentum and its modulus is written as \textit{P}. Therefore, we have obtained a model-independent integral formula for the chiral susceptibility at finite temperature and vanishing chemical potential.

\subsection{Formalism of the fermion number susceptibility}
In addition to the above chiral phase transition, as temperature and/or chemical potential increase, strongly interacting matter will also experience a phase transition from the confinement phase (where the degree of freedom for system is hadron and thus the baryon number is an integer) to the deconfinement phase (in which the degree of freedom is quark and gluon, and so the baryon number is a fraction).  The fluctuation of fermion number is theoretically constructed from measurement of the fermion number susceptibility, i.e the response of fermion number density to an infinitesimal change in the chemical potential.

The fermion number susceptibility is defined as
\begin{eqnarray}
\chi^{n}=\frac{\partial\rho(\mu)}{\partial\mu}\bigg|_{\mu=0}=\frac{T}{V}\frac{\partial^{2}\ln\mathcal{Z}}{\partial \mu^{2}}\bigg|_{\mu=0},\label{eq9}
\end{eqnarray}
where $\rho$ represents the fermion number density, and $\mu$ is the chemical potential of the fermion.

Meanwhile, the fermion number density can be expressed as~\cite{PhysRevD.78.054001}
\begin{eqnarray}
\rho(\mu)=-\int\frac{\textrm{d}^{3}p}{(2\pi)^{3}}\textrm{Tr}[\textit{S}(\mu,p)\gamma^{3}],\label{eq10}
\end{eqnarray}
where $\textit{S}(\mu,p)$ is the dressed fermion propagator at finite current mass and chemical potential.

Substituting Eq. (10) into Eq. (9) and using the identity
\begin{eqnarray}
\frac{\partial S(\mu,p)}{\partial \mu}=-S(\mu,p)\frac{\partial S^{-1}(\mu,p)}{\partial \mu}S(\mu,p),\nonumber\\\label{eq11}
\end{eqnarray}
one easily arrive at
\begin{eqnarray}
\chi^{n}=\int\frac{\textrm{d}^{3}p}{(2\pi)^{3}}\textrm{Tr}[S(\mu,p)\frac{\partial S^{-1}(\mu,p)}{\partial \mu}S(\mu,p)\gamma^{3}]\bigg|_{\mu=0},\nonumber\\\label{eq12}
\end{eqnarray}

In the same way as mentioned above, we consider $\mathcal{A}^{\nu}$ as a constant background vector field coupled to the fermion fields in the manner $\bar\psi\gamma^{\nu}\psi\mathcal{A}^{\nu}$. Then \textit{S}($\mathcal{A},\textit{p}$) is the dressed fermion propagator in the presence of this background field and the derivative of its inverse with respect to $\mathcal{A}^{\nu}$ gives the so-called 1PI dressed vector vertex
\begin{eqnarray}
\Gamma^{\nu}(\mathcal{A},0,p)=-\frac{\partial S^{-1}(\mathcal{A},p)}{\partial \mathcal{A}^{\nu}},\label{eq13}
\end{eqnarray}
where $\textit{p}$ is the relative momentum and the total momentum of this vertex still vanishes due to the constant background field. Putting $\mathcal{A}^{\nu}=\delta^{\nu3}\mu$ in Eq. (13), we obtain
\begin{eqnarray}
\Gamma^{3}(\mu,0,p)=-\frac{\partial S^{-1}(\mu,p)}{\partial \mu},\label{eq14}
\end{eqnarray}

Combining Eq. (12) with Eq. (14) gives
\begin{eqnarray}
\chi^{n}=-\int\frac{\textrm{d}^{3}p}{(2\pi)^{3}}\textrm{Tr}[S(\mu,p)\Gamma^{3}(\mu,0,p)S(\mu,p)\gamma^{3}]\bigg|_{\mu=0},\nonumber\\\label{eq15}
\end{eqnarray}
thus, we have also got the expression for the fermion number susceptibility, in term of dressed fermion propagator and dressed vector vertex which can be calculated by means of DSEs-BSE approach at zero temperature and chemical potential.

Finally, similar to Eq. (8), the above equation can be generalized to the finite temperature version
\begin{eqnarray}
\chi^{n}(T)&=&-\textit{T}\sum_{n}\int\frac{\textrm{d}^{2}P}{(2\pi)^{2}}\textrm{Tr}[S(\mu,p_{n})\Gamma^{3}(\mu,0,p_{n})\nonumber\\
&&\times S(\mu,p_{n})\gamma^{3}]\bigg|_{\mu=0},\label{eq16}
\end{eqnarray}
therefore, we have obtained a model-independent integral formula for the fermion number susceptibility at finite temperature and vanishing chemical potential.

\subsection{Formalism of the thermal susceptibility}
The chiral condensate is commonly used to characterize the chiral phase transition of strongly interacting matter. In the case of finite temperature, it is a function of temperature and shows some singular behaviors near the phase transition point. In the issue, the thermal susceptibility, that is, the response of the chiral condensate to infinitesimal change of temperature, has attracted quite a few interests over the years. Some works in Ref.~\cite{Phys.Rev.C.58.1758,Phys.Rev.D.84.074020,Phys.Rev.D.85.034026} showed that the peak of the thermal susceptibility should be a crucial character of chiral phase transition.

As mentioned above, the thermal susceptibility is defined as
\begin{eqnarray}
\chi^{T}(T)=\frac{\partial\langle\bar\psi\psi\rangle}{\partial T}=-\frac{1}{V}\frac{\partial^{2}(T\ln\mathcal{Z})}{\partial T \partial m},\label{eq17}
\end{eqnarray}
where $\langle\bar\psi\psi\rangle$ denotes the finite-temperature fermion chiral condensate at finite current mass, which can be obtained from Eq. (2) by replacing the integration over the third component of momentum with summation over Matsubara frequencies
\begin{eqnarray}
-\langle\bar\psi\psi\rangle=\textit{T}\sum_{n}\int\frac{\textrm{d}^2P}{(2\pi)^2}\textrm{Tr}[S(\textit{m},p_{n})].\label{eq18}
\end{eqnarray}

Substituting Eq. (18) into Eq. (17), the thermal susceptibility can be expressed as
\begin{eqnarray}
\chi^{T}(T)=-\textit{T}\sum_{n}\int\frac{\textrm{d}^2P}{(2\pi)^2}\textrm{Tr}[\frac{S(\textit{m},p_{n})}{T}+\frac{\partial S(\textit{m},p_{n})}{\partial T}],\nonumber\\
\label{eq19}
\end{eqnarray}
therefore, once the finite-temperature dressed fermion propagator is known, one can calculate the thermal susceptibility.

\subsection{Formalism of the staggered spin susceptibility}
Because of its success in interpreting the existence of antiferromagnetic correlation in underdoped cuprates, the U(1) gauge fluctuation effect has aroused great interest and extensive attention both in theory and experiment in recent years. The response corresponding to this fluctuation is the staggered spin susceptibility which can be directly measured in experiments and so provides an ideal tool to probe the characteristics of strongly correlated system. In recent work, based on functional analysis, the general formula for the staggered spin susceptibility was given in term of dressed fermion propagator and dressed pseudoscalar vertex, and thus it can be calculated within the framework of DSEs-BSE approach.

The general expression for the low-energy behavior of the regularized staggered spin susceptibility was given by Ref.~\cite{Phys.Rev.D.87.116008}
\begin{eqnarray}
\chi^{s}=\int\frac{\textrm{d}^{3}p}{(2\pi)^{3}}\textrm{Tr}[S(p)\Gamma_{\textrm{p}}(p)S(p)-S_{0}(p)\textbf{1}S_{0}(p)],\label{eq20}
\end{eqnarray}
where the notation $\Gamma_{\textrm{p}}$ represents pseudoscalar vertex that satisfies the corresponding inhomogeneous BSE.

The corresponding finite temperature version of the staggered spin susceptibility can be obtained
\begin{eqnarray}
\chi^{s}(T)&=&\textit{T}\sum_{n}\int\frac{\textrm{d}^2P}{(2\pi)^2}\textrm{Tr}[S(p_{n})\Gamma_{\textrm{p}}(p_{n})S(p_{n})\nonumber\\
&&-S_{0}(p_{n})\textbf{1}S_{0}(p_{n})],\label{eq21}
\end{eqnarray}
where $\Gamma_{\textrm{p}}$ satisfies the finite temperature version of the inhomogeneous BSE.

\section{NUMERICAL RESULTS}

\subsection{Dyson-Schwinger equations in QED$_3$}
Given the chiral symmetry and parity transformation, we will employ the four-dimension matrix representation and four-component spinors as in four space-time dimensions. In Euclidean space, the Lagrangian density of QED$_3$ with $\textit{N}_{f}$ massless fermion flavors reads
\begin{eqnarray}
\mathcal{L}=\sum _{f=1}^{N_{f}} \bar{\psi }_f(-{\not\!\partial}-m +ie{\not\!A})\psi _f-\frac{1}{4}F_{\mu \nu }^2-\frac{1}{2\xi}(\partial\cdot A)^2,\nonumber\\\label{eq22}
\end{eqnarray}
where the subscript $\textit{f}$ is a flavour label; $\textit{f}$ = 1, 2, ..., $N_{f}$ for a theory with $N_{f}$ distinct types or flavours of electrically active fermions. We will only work with one flavor in the present paper. Using this Lagrangian density, one can derive in the standard way, for instance through functional analysis, the DSEs for propagators.

For the fermion propagator, we obtain the finite temperature version of DSEs
\begin{eqnarray}
S^{-1}(m,p_{n})=S_{0}^{-1}(m,p_{n}) + \Sigma(m,p_{n}),\label{eq23}
\end{eqnarray}
\begin{eqnarray}
\Sigma(m,p_{n})=\textit{T}\sum_{n}\int\frac{\textrm{d}^{2}K}{(2\pi)^{2}}\gamma_{\mu}S(m,k_{n})\Gamma_{\nu}(p_{n},k_{n})D_{\mu\nu}(q_{n}),\nonumber\\\label{eq24}
\end{eqnarray}
where $S_{0}^{-1}=i\vec{\gamma}\cdot\vec{p}+i\gamma_{3}\omega_{n}+m$ is just inverse of the free fermion propagator, $\Sigma$ is the fermion self-energy, $\Gamma_{\nu}$ is the full fermion-boson vertex, and $D_{\mu\nu}$ is the dressed boson propagator.

Other than zero temperature, the O(3) symmetry of the system reduces to O(2), and based on the Lorentz structure analysis, the inverse of fermion propagator can be written as
\begin{eqnarray}
S^{-1}(\textit{m},p_{n})&=&\textit{i}\vec{\gamma}\cdot\vec{\textit{p}}\textit{A}_{\parallel}(m,p_{n})+\textit{i}\gamma_{3}\omega_{n}\textit{A}_3(m,p_{n})\nonumber\\
&&+\textit{B}(m,p_{n}),\label{eq25}
\end{eqnarray}
where $\textit{A}_{\parallel}$ and $\textit{A}_3$ are familiar wave function renormalization factors; $B$ is fermion self-energy function, and a tensor term proportional to $\sigma_{\mu\nu}$ is ruled out by $\mathcal{PT}$ invariance.

At zero temperature, the results in Ref.~\cite{Phys.Rev.D.70.073007} show that when the 1/N order contribution to the renormalization factor is included, the critical number of fermion flavors take almost the same value as the case where $A=1$. At finite temperature, the comparison of studies in Ref.~\cite{Phys.Lett.B.166.163,Phys.Rev.D.50.1068} also suggests that the 1/N order contribution to the factor only changes the results slightly. So we expect that the 1/N order contribution to $\textit{A}_{\parallel}$ and $\textit{A}_3$ is not important and we will take $\textit{A}_{\parallel}$=$\textit{A}_3$=1 in the present paper. In addition, the conclusions in Ref.~\cite{Phys.Rev.D.67.076005} indicate that by summing over the frequency modes and taking suitable simplifications, the qualitative aspects of the result obtained under the zero frequency approximation for fermion self-energy don't undergo significant changes. From this, we will also ignore the frequency dependence of the self-energy.

For the boson propagator, we will follow Ref.~\cite{Phys.Lett.B.166.163} in retaining only the $\mu=\nu=3$ component of the boson propagator, and ignore all but the zero-frequency component, that is to say, we employ the boson propagator
\begin{eqnarray}
D_{\mu\nu}(m,T,Q)=\frac{2\delta_{\mu3}\delta_{\nu3}}{Q^{2}+\Pi(m,T,Q)},\label{eq26}
\end{eqnarray}
where $Q^2=\vec{q}^{2}=(\vec{p}-\vec{k})^2$. The zero frequency boson polarization with current mass reads
\begin{eqnarray}
\Pi(m,T,Q)&=&\frac{T}{\pi}\int^{1}_{0}\textrm{d}x\bigg\{\textrm{ln}(4\textrm{cosh}^{2}\frac{\sqrt{m^{2}+x(1-x)Q^{2}}}{2T})\nonumber\\
&&-\frac{m^{2}\textrm{tanh}\frac{\sqrt{m^{2}+x(1-x)Q^{2}}}{2T}}{T\sqrt{m^{2}+x(1-x)Q^{2}}}\bigg\},\label{eq27}
\end{eqnarray}

Substituting Eq. (24), Eq. (25) and Eq. (26) into Eq. (23), we can immediately obtain (to be concise, hereafter we use $B$ to represent $B(m,T,P^{2})$, and $\Pi$ to represent $\Pi(m,T,Q)$ in the right sides of equations)
\begin{eqnarray}
B(m,T,P^{2})=m+\textit{T}\sum_{n}\int\frac{\textrm{d}^{2}K}{(2\pi)^{2}}\frac{B/(Q^{2}+\Pi)}{(\omega_{n}^{2}+K^{2}+B^{2})},\nonumber\\\label{eq28}
\end{eqnarray}
with the help of the identity
\begin{eqnarray}
\sum_{n}\frac{1}{\omega_{n}^{2}+x^{2}}=\frac{\textrm{tanh}\frac{x}{2T}}{2xT},\label{eq29}
\end{eqnarray}
the above equation can be reduced to
\begin{eqnarray}
B(m,T,P^{2})=m+\int\frac{\textrm{d}^{2}K}{(2\pi)^{2}}\frac{B\textrm{tanh}\frac{\sqrt{K^{2}+B^{2}}}{2T}}{\sqrt{K^{2}+B^{2}}(Q^{2}+\Pi)}.\nonumber\\\label{eq30}
\end{eqnarray}

\subsection{Chiral symmetry breaking-restoration phase transition in the chiral limit}
In this section, within the framework of DSEs approach, we will investigate the behavior of the four susceptibilities with varying temperature. All equations involved in the section are in the chiral limit.

\subsubsection{The chiral susceptibility $\chi^{c}$}
For the chiral susceptibility, substituting Eq. (5) and Eq. (25) into Eq. (8) gives
\begin{eqnarray}
\chi^{c}_{r}(T)&=&2\int\frac{\textrm{d}^{2}\textit{P}}{(2\pi)^{2}}\bigg\{B_{m}[\frac{P^{2}\textrm{tanh}\frac{\sqrt{P^{2}+B^{2}}}{2T}}{(\sqrt{P^{2}+B^{2}})^{3}}\nonumber\\
&&+\frac{B^{2}\textrm{sech}^{2}\frac{\sqrt{P^{2}+B^{2}}}{2T}}{2T(P^{2}+B^{2})}]-\frac{1}{P}\textrm{tanh}\frac{P}{2T}\bigg\},\label{eq31}
\end{eqnarray}
where $B$ is the function that satisfies Eq. (30) in the absence of current mass. The function $B_{m}$ is just the derivative of $B$ with respect to current mass and can be written as
\begin{eqnarray}
B_{m}&=&1+\int\frac{\textrm{d}^{2}K}{(2\pi)^{2}}B_{m}\bigg\{\frac{K^{2}\textrm{tanh}\frac{\sqrt{K^{2}+B^{2}}}{2T}}{(\sqrt{K^{2}+B^{2}})^{3}}\nonumber\\
&&+\frac{B^{2}\textrm{sech}^{2}\frac{\sqrt{K^{2}+B^{2}}}{2T}}{2T(K^{2}+B^{2})}\bigg\}\frac{1}{Q^{2}+\Pi},\label{eq32}
\end{eqnarray}

\begin{figure}
\includegraphics[width=0.47\textwidth]{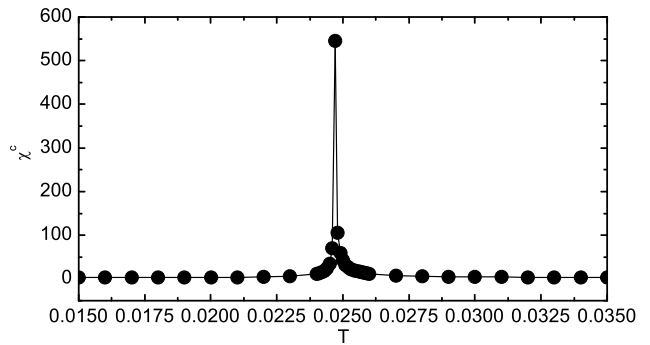}
\caption{Dependence of the chiral susceptibility $\chi^{c}$ on temperature \textit{T}.}\label{fig1}
\end{figure}

Using the iterative method for the above two equations, we can immediately arrive at the typical behavior of them, and thus the chiral susceptibility. As a result, we plot the behavior of the chiral susceptibility with varying temperature in Fig. \ref{fig1}~\footnote{In the studies within QED$_3$, the coupling constant $\alpha=e^2$ has dimension one and provides us with a mass scale. Accordingly, a kind of natural unit $e^2=1$ is often used (for example, see Ref.~\cite{PhysRevD.86.045020}). For simplicity, in this paper the temperature and the mass are both measured in this unit.}.

From Fig. 1, it can be seen that the chiral susceptibility exhibits a very narrow, pronounced, and, in fact, divergent peak at the critical temperature $T_{c}=2.47\times10^{-2}$, which is a typical characteristic of the second order phase transition. This conclusion is in good agreement with the result based on the continuum model studies of two-flavor QCD~\cite{Phys.Rev.D.77.076008}.

\subsubsection{The fermion number susceptibility $\chi^{n}$}
For the fermion number susceptibility, substituting Eq. (14) and Eq. (25) into Eq. (16), and employing the similar approximation in Ref.~\cite{Phys.Rev.D.86.065002} for the dressed vector vertex, we can obtain
\begin{eqnarray}
\chi^{n}(T)&=&\frac{4}{T}\int\frac{\textrm{d}^{2}P}{(2\pi)^{2}}\frac{\exp\frac{\sqrt{P^{2}+B^{2}}}{T}}{(\exp\frac{\sqrt{P^{2}+B^{2}}}{T}+1)^{2}},\label{eq33}
\end{eqnarray}

\begin{figure}
\includegraphics[width=0.47\textwidth]{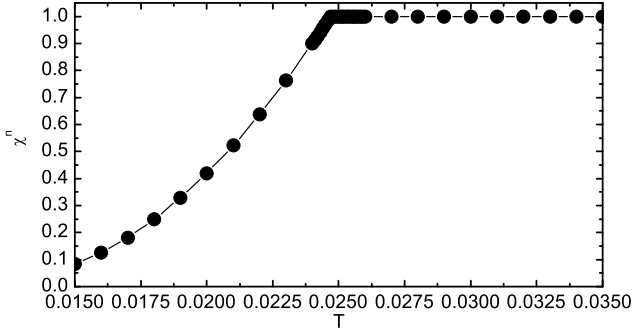}
\caption{Dependence of the fermion number susceptibility $\chi^{n}/\chi^{n}_{f}$ on temperature \textit{T}, where $\chi^{n}_{f}$ is the free fermion number susceptibility at finite temperature.}\label{fig2}
\end{figure}

Once the fermion self-energy function is obtained, we can calculate the fermion number susceptibility. The behavior of $\chi^{n}(T)$ is shown in Fig. \ref{fig2}.

As is shown in Fig. 2, where $\chi^{n}$ is normalized by the free fermion number susceptibility $\chi^{n}_{f}$ and is hence dimensionless, the fermion number susceptibility rises as temperature increases, then shows an apparent inflexion at the critical temperature $T_{c}=2.47\times10^{-2}$, and finally is almost constant.

\subsubsection{The thermal susceptibility $\chi^{T}$}
\begin{figure}
\includegraphics[width=0.47\textwidth]{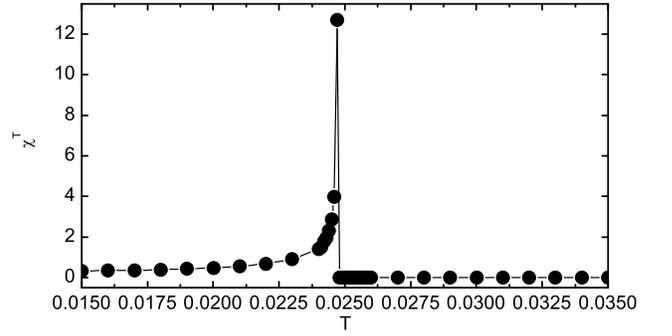}
\caption{Dependence of the thermal susceptibility $\chi^{T}$ on temperature \textit{T}.}\label{fig3}
\end{figure}

For the thermal susceptibility, substituting Eq. (25) into Eq. (19), then the latter can be reduced to
\begin{eqnarray}
\chi^{T}(T)&=&2\int\frac{\textrm{d}^{2}P}{(2\pi)^{2}}\bigg\{B_{T}[\frac{P^{2}\textrm{tanh}\frac{\sqrt{P^{2}+B^{2}}}{2T}}{(\sqrt{P^{2}+B^{2}})^{3}}\nonumber\\
&&+\frac{B^{2}\textrm{sech}^{2}\frac{\sqrt{P^{2}+B^{2}}}{2T}}{2T(P^{2}+B^{2})}]-\frac{B\textrm{sech}^{2}\frac{\sqrt{P^{2}+B^{2}}}{2T}}{2T^{2}}\bigg\},\nonumber\\\label{eq34}
\end{eqnarray}
where the function $B_{T}$ represents the derivative of fermion self-energy function with respect to temperature, and can be expressed as
\begin{eqnarray}
B_{T}&=&\int\frac{\textrm{d}^{2}K}{(2\pi)^{2}}\bigg\{B_{T}[\frac{K^{2}\textrm{tanh}\frac{\sqrt{K^{2}+B^{2}}}{2T}}{(\sqrt{K^{2}+B^{2}})^{3}}\nonumber\\
&&+\frac{B^{2}\textrm{sech}^{2}\frac{\sqrt{K^{2}+B^{2}}}{2T}}{2T(K^{2}+B^{2})}]-\frac{B\textrm{sech}^{2}\frac{\sqrt{K^{2}+B^{2}}}{2T}}{2T^{2}}\nonumber\\
&&-\frac{B\Pi_{T}\textrm{tanh}\frac{\sqrt{K^{2}+B^{2}}}{2T}}{\sqrt{K^{2}+B^{2}}(Q^{2}+\Pi)}\bigg\}\frac{1}{Q^{2}+\Pi},\label{eq35}
\end{eqnarray}
while the function $\Pi_{T}$ denotes the derivative of boson polarization function with respect to the temperature, which is written as
\begin{eqnarray}
\Pi_{T}&=&\frac{1}{\pi}\int^{1}_{0}\bigg\{\ln(4\textrm{cosh}^{2}\frac{\sqrt{x(1-x)Q^{2}}}{2T})\nonumber\\
&&-\frac{\sqrt{x(1-x)Q^{2}}\textrm{tanh}\frac{\sqrt{x(1-x)Q^{2}}}{2T}}{T}\bigg\},\label{eq36}
\end{eqnarray}
from the two functions $B$ and $B_{T}$, one can obtain the dependence of the thermal susceptibility on temperature. As a result, the behavior of this susceptibility is plotted in Fig. \ref{fig3}.

From Fig. 3, it is found that the thermal susceptibility rises with temperature increasing, then reaches its maximum at the critical temperature $T_{c}=2.47\times10^{-2}$, and vanishes when temperature is above $T_{c}$

\subsubsection{The staggered spin susceptibility $\chi^{s}$}
For the staggered spin susceptibility, we focus on its low-energy behavior, and so the dressed pseudoscalar vertex $\Gamma_{\textrm{p}}$ can be written as
\begin{eqnarray}
\Gamma_{\textrm{p}}(p_{n})=\gamma_{5}C(p_{n}^{2})+i\vec{\gamma}\cdot\vec{p}\gamma_{5}D_{\parallel}(p_{n}^{2})+i\gamma_{3}w_{n}\gamma_{5}D_{3}(p_{n}^{2}),\nonumber\\
\label{eq37}
\end{eqnarray}
Substituting Eq. (37) into Eq. (21), we immediately arrive at
\begin{eqnarray}
\chi^{s}(T)=2\int\frac{\textrm{d}^2P}{(2\pi)^2}(\frac{C\textrm{tanh}\frac{\sqrt{P^{2}+B^{2}}}{2T}}{\sqrt{P^{2}+B^{2}}}-\frac{\textrm{tanh}\frac{P}{2T}}{P}).\label{eq38}
\end{eqnarray}

Meanwhile, the dressed pseudoscalar vertex satisfies the finite temperature version of the inhomogeneous BSE
\begin{eqnarray}
\Gamma_{\textrm{p}}(p_{n})&=&\gamma_{5}-\textit{T}\sum_{n}\int\frac{\textrm{d}^{2}K}{(2\pi)^{2}}\gamma_{\mu}S(k_{n})\Gamma_{\textrm{p}}(k_{n})S(k_{n})\gamma_{\nu}\nonumber\\
&&\times D_{\mu\nu}(q_{n}),\label{eq39}
\end{eqnarray}
Substituting Eq. (37) into Eq. (39) gives
\begin{eqnarray}
C(P^{2})=1+\int\frac{\textrm{d}^2K}{(2\pi)^2}\frac{C\textrm{tanh}\frac{\sqrt{K^{2}+B^{2}}}{2T}}{\sqrt{K^{2}+B^{2}}(Q^{2}+\Pi)}.\label{eq40}
\end{eqnarray}

\begin{figure}
\includegraphics[width=0.47\textwidth]{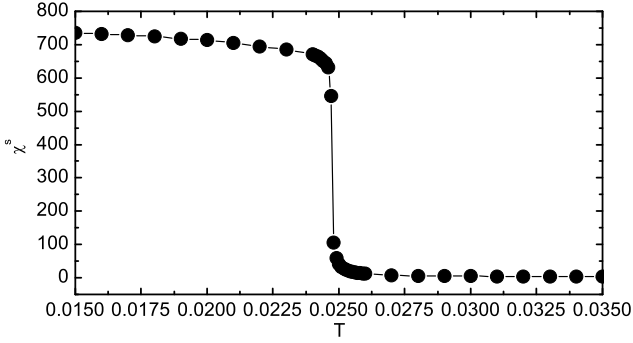}
\caption{Dependence of the staggered spin susceptibility $\chi^{s}$ on temperature \textit{T}.}\label{fig4}
\end{figure}

By solving the functions $B$ and $C$, we can obtain the staggered spin susceptibility with a range of temperature, and the results are shown in Fig. \ref{fig4}.

As can be seen from Fig. 4, the staggered spin susceptibility decreases with temperature increasing, then drops rapidly at the critical temperature $T_{c}=2.47\times10^{-2}$, and is almost constant when temperature is higher.

From above, we may safely draw the conclusion that these four susceptibilities exhibit the singular behaviors at the same critical temperature $T_{c}=2.47\times10^{-2}$, where the chiral phase transition occurs. In addition, the chiral susceptibility exhibits a divergent peak at the critical temperature, which is a typical characteristic of the second order phase transition.

\subsection{Chiral symmetry breaking-restoration phase transition beyond the chiral limit}
In this section, we will recalculate the four susceptibilities to analyze the effect of current mass on the chiral phase transition driven by temperature.

\subsubsection{The chiral susceptibility $\chi^{c}$}
Regarding the chiral susceptibility, following the same step as Eq. (31), the chiral susceptibility beyond the chiral limit is obtained
\begin{eqnarray}
\chi^{c}_{r}(m,T)&=&2\int\frac{\textrm{d}^{2}\textit{P}}{(2\pi)^{2}}\bigg\{B_{m}[\frac{P^{2}\textrm{tanh}\frac{\sqrt{P^{2}+B^{2}}}{2T}}{(\sqrt{P^{2}+B^{2}})^{3}}\nonumber\\
&&+\frac{B^{2}\textrm{sech}^{2}\frac{\sqrt{P^{2}+B^{2}}}{2T}}{2T(P^{2}+B^{2})}]-[\frac{P^{2}\textrm{tanh}\frac{\sqrt{P^{2}+m^{2}}}{2T}}{(\sqrt{P^{2}+m^{2}})^{3}}\nonumber\\
&&+\frac{m^{2}\textrm{sech}^{2}\frac{\sqrt{P^{2}+m^{2}}}{2T}}{2T(P^{2}+m^{2})}]\bigg\},\label{eq41}
\end{eqnarray}
where the self-energy function $B$ satisfies Eq. (30) and thus its derivative with respect to current mass can be expressed as
\begin{eqnarray}
B_{m}&=&\int\frac{\textrm{d}^{2}K}{(2\pi)^{2}}\bigg\{B_{m}[\frac{K^{2}\textrm{tanh}\frac{\sqrt{K^{2}+B^{2}}}{2T}}{(\sqrt{K^{2}+B^{2}})^{3}}
+\frac{B^{2}\textrm{sech}^{2}\frac{\sqrt{K^{2}+B^{2}}}{2T}}{2T(K^{2}+B^{2})}]\nonumber\\
&&-\frac{B\Pi_{m}\textrm{tanh}\frac{\sqrt{K^{2}+B^{2}}}{2T}}{\sqrt{K^{2}+B^{2}}(Q^{2}+\Pi)}\bigg\}\frac{1}{Q^{2}+\Pi}+1,\label{eq42}
\end{eqnarray}
the function $\Pi_{m}$ is the derivative of polarization function with respect to current mass, which satisfies
\begin{eqnarray}
\Pi_{m}&=&-\frac{m}{\pi}\int^{1}_{0}[\frac{x(1-x)Q^{2}\textrm{tanh}\frac{\sqrt{m^{2}+x(1-x)Q^{2}}}{2T}}{(\sqrt{m^{2}+x(1-x)Q^{2}})^{3}}\nonumber\\
&&+\frac{m^{2}\textrm{sech}^{2}\frac{\sqrt{m^{2}+x(1-x)Q^{2}}}{2T}}{2T(m^{2}+x(1-x)Q^{2})}],\label{eq43}
\end{eqnarray}

\begin{figure}
\includegraphics[width=0.47\textwidth]{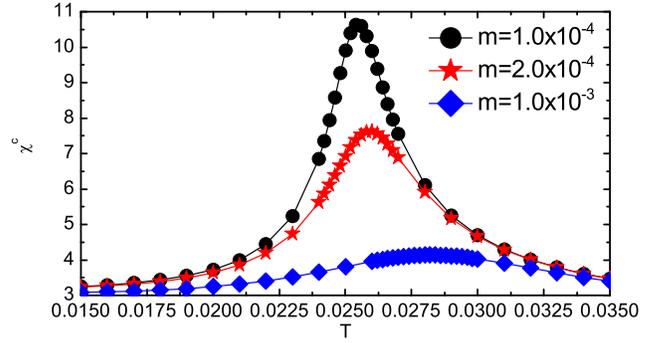}
\caption{Dependence of the chiral susceptibility $\chi^{c}$ on temperature \textit{T} for several \textit{m}.}\label{fig5}
\end{figure}

From Eq. (30), Eq. (41) and Eq. (42), we can obtain the dependence of the chiral susceptibility on temperature and current mass. The behaviors of $\chi^{c}$ with regard to $T$ for several $\textit{m}$ are plotted in Fig. \ref{fig5}.

As is shown in Fig. 5, the chiral susceptibility exhibits a quite different behavior in the presence of current mass. The peak of the chiral susceptibility becomes
not so sharp and pronounced as in the chiral limit and its height is greatly suppressed and evidently finite, which is a typical character of a crossover. In addition, with the current mass increasing, the critical temperature where the chiral susceptibility takes its maximum also rises, but the value of the peak falls monotonously.

\subsubsection{The fermion number susceptibility $\chi^{n}$}
Similar to Eq. (33), the fermion number susceptibility beyond the chiral limit can be written as
\begin{eqnarray}
\chi^{n}(m,T)&=&\frac{4}{T}\int\frac{\textrm{d}^{2}P}{(2\pi)^{2}}\frac{\exp\frac{\sqrt{P^{2}+B^{2}}}{T}}{(\exp\frac{\sqrt{P^{2}+B^{2}}}{T}+1)^{2}},\label{eq44}
\end{eqnarray}
where the fermion self-energy function $B$ satisfies Eq. (30).

\begin{figure}
\includegraphics[width=0.47\textwidth]{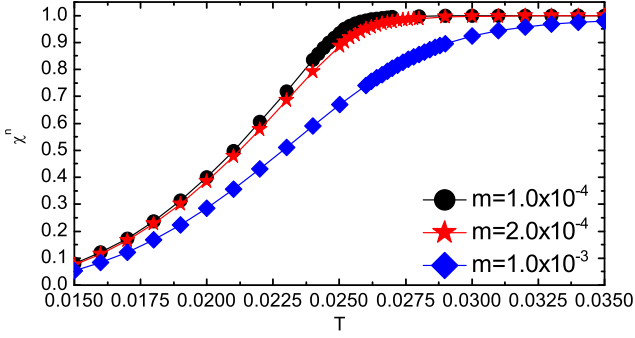}
\caption{Dependence of the fermion number susceptibility $\chi^{n}/\chi^{n}_{f}$ on temperature \textit{T} for several \textit{m}, where $\chi^{n}_{f}$ is the free fermion number susceptibility at finite temperature and current mass of the fermion.}\label{fig6}
\end{figure}

From Eq. (27), Eq. (30) and Eq. (44), the dependence of the fermion number susceptibility on temperature and current mass is immediately obtained. We show the behaviors of $\chi^{n}$ with respect to $T$ for several $\textit{m}$ in Fig. \ref{fig6}.

In Fig. 6, the fermion number susceptibility at finite current mass also reveals a different picture from the case of chiral limit. As temperature increases, the fermion number susceptibility also rises monotonously and almost keeps a constant at last. It is noteworthy that $\chi^{n}$ becomes smooth and no singular behaviors emerge in a range of temperature, which may show that the fermion number susceptibility can not describe the crossover properties of the system well. It is consistent with the result obtained using the NJL model~\cite{Phys.Rev.D.88.114019} .

\subsubsection{The thermal susceptibility $\chi^{T}$}
For the thermal susceptibility, analogizing with Eq. (34), we finally arrive at
\begin{eqnarray}
\chi^{T}(m,T)&=&2\int\frac{\textrm{d}^{2}P}{(2\pi)^{2}}\bigg\{B_{T}[\frac{P^{2}\textrm{tanh}\frac{\sqrt{P^{2}+B^{2}}}{2T}}{(\sqrt{P^{2}+B^{2}})^{3}}\nonumber\\
&&+\frac{B^{2}\textrm{sech}^{2}\frac{\sqrt{P^{2}+B^{2}}}{2T}}{2T(P^{2}+B^{2})}]-\frac{B\textrm{sech}^{2}\frac{\sqrt{P^{2}+B^{2}}}{2T}}{2T^{2}}\bigg\},\nonumber\\\label{eq45}
\end{eqnarray}
where the two functions $B$ and $B_{T}$ satisfy, separately, Eq. (30) and Eq. (35). The function $\Pi_{T}$ involved here is a little different from Eq. (36), and is written as
\begin{eqnarray}
\Pi_{T}&=&\frac{1}{\pi}\int^{1}_{0}\bigg\{\ln(4\textrm{cosh}^{2}\frac{\sqrt{m^{2}+x(1-x)Q^{2}}}{2T})\nonumber\\
&&-\frac{\sqrt{m^{2}+x(1-x)Q^{2}}\textrm{tanh}\frac{\sqrt{m^{2}+x(1-x)Q^{2}}}{2T}}{T}\nonumber\\
&&+\frac{m^{2}\textrm{sech}^{2}\frac{\sqrt{m^{2}+x(1-x)Q^{2}}}{2T}}{2T^{2}}\bigg\},\label{eq46}
\end{eqnarray}

\begin{figure}
\includegraphics[width=0.47\textwidth]{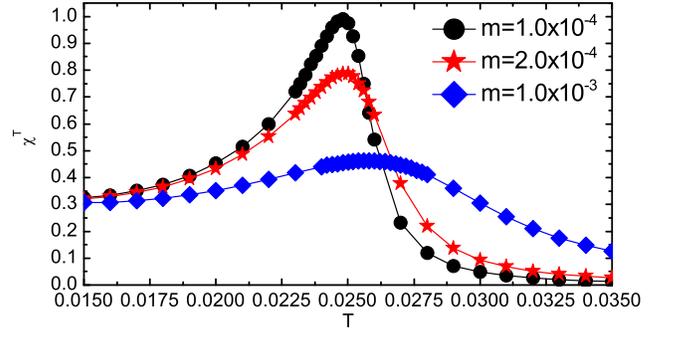}
\caption{Dependence of the thermal susceptibility $\chi^{T}$ on temperature \textit{T} for several \textit{m}.}\label{fig7}
\end{figure}

According to the equation above, we obtain the dependence of the thermal susceptibility on temperature and current masses. As a result, the behaviors of $\chi^{T}$ with regard to temperature for several current masses are plotted in Fig. \ref{fig7}.

From Fig. 7, we can evidently see that the thermal susceptibility shows an apparent peak. With temperature increasing, the thermal susceptibility also rises, then takes its maximum and decreases slowly when the temperature is higher. Similar to the chiral susceptibility, as current mass increases, the critical temperature at which the thermal susceptibility is maximal rises, while the value of the peak falls slowly.

\subsubsection{The staggered spin susceptibility $\chi^{s}$}
Following the same step as Eq. (38), the staggered spin susceptibility in the presence of current mass is expressed as
\begin{eqnarray}
\chi^{s}(m,T)&=&2\int\frac{\textrm{d}^2p}{(2\pi)^2}\bigg\{\frac{C\textrm{tanh}\frac{\sqrt{P^{2}+B^{2}}}{2T}}{\sqrt{P^{2}+B^{2}}}\nonumber\\
&&-[\frac{P^{2}\textrm{tanh}\frac{\sqrt{P^{2}+m^{2}}}{2T}}{(\sqrt{P^{2}+m^{2}})^{3}}+\frac{m^{2}\textrm{sech}^{2}\frac{\sqrt{P^{2}+m^{2}}}{2T}}{2T(P^{2}+m^{2})}]\bigg\},\nonumber\\
\label{eq47}
\end{eqnarray}

\begin{figure}
\includegraphics[width=0.47\textwidth]{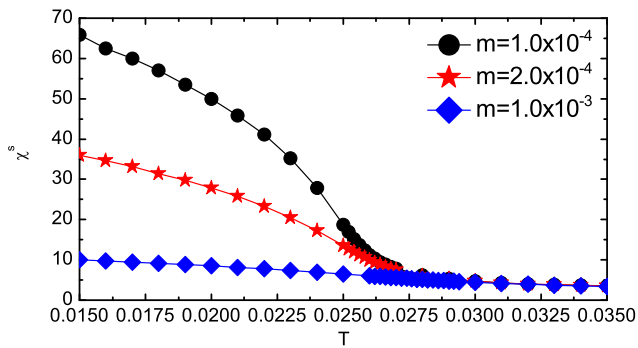}
\caption{Dependence of the staggered spin susceptibility $\chi^{s}$ on temperature \textit{T} for several \textit{m}.}\label{fig8}
\end{figure}

By solving Eq. (30) and Eq. (40), the dependence of the staggered spin susceptibility on temperature and current mass can be obtained. We plot the behaviors of $\chi^{s}$ with respect to temperature at several current masses in Fig. \ref{fig8}.

From Fig. 8, it can be seen that the staggered spin susceptibility in this case shows a quite different picture from the chiral limit case. With the temperature increasing, the staggered spin susceptibility decreases slowly and is almost a constant in the end. Similar to the fermion number susceptibility, the staggered spin susceptibility is smooth and no singular behaviors occur in the temperature range we studied.

From what we have mentioned above, we can conclude that the four susceptibilities in the presence of current mass have intrinsic differences from the cases of chiral limit. Both the chiral and the thermal susceptibilities reveal an apparent but not divergent peak signalling a typical crossover behavior, meanwhile the critical temperatures from these two susceptibilities are different, which shows that to define a critical region instead of an exclusive point for crossover might be a more suitable choice. Moreover, the fermion number and the staggered spin susceptibilities are smooth with varying temperature and no singular behaviors arise, this may mean that these two susceptibilities can not describe the crossover properties of the system well.

\section{summary and conclusions}
In this paper, we study the relations of the four different susceptibilities (viz., the chiral susceptibility, the fermion number susceptibility, the thermal susceptibility and the staggered spin susceptibility) both in and beyond the chiral limit. We first give the general integral formula for the four different susceptibilities by means of the external field method, and then investigate the temperature dependence of them in the framework of DSEs.

Our model study reveals that, in the chiral limit, the four susceptibilities give the same critical temperature $T_{c}=2.47\times10^{-2}$, where the chiral phase transition occurs. In addition, the chiral susceptibility shows that this is a second order phase transition at finite temperature and vanishing chemical potential. On the other hand, in the presence of current mass, the results are quite different: the critical temperatures from the chiral and the thermal susceptibilities are different, which shows that to define a critical region instead of an exclusive point for crossover might be a more suitable choice. In addition, both the fermion number and the staggered spin susceptibilities have no singular behaviors any more, this may mean that they are no longer available to describe the crossover properties of the system.

Of course, the model adopted in this work is schematic, to further confirm these observations, we need to study this problem in some more realistic models in the future.

\acknowledgments

This work is supported in part by the National Natural Science Foundation of China (under Grant 11275097, 10935001, 11274166, 11247219 and 11075075), the National Basic Research Program of China (under Grant 2012CB921504), the Research Fund for the Doctoral Program of Higher Education (under Grant No 2012009111002) and the National Natural Science Foundation of Jiangsu Province of China (under Grant BK20130078).

\bibliographystyle{apsrev4-1}
\bibliography{DT11264}

\end{document}